\documentstyle[12pt,epsf]{article} 

\def\lsim{\mathrel{\lower2.5pt\vbox{\lineskip=0pt\baselineskip=0pt 
           \hbox{$<$}\hbox{$\sim$}}}} 
\def\gsim{\mathrel{\lower2.5pt\vbox{\lineskip=0pt\baselineskip=0pt 
           \hbox{$>$}\hbox{$\sim$}}}}

\def\Ap{A^{\prime}}

\def\App{A^{\prime\prime}}

\def\k{\kappa}
\def\p{\partial}
\def\R{{\cal R}}
\def\d{\delta}

\def\L{\Lambda}

\def\P{\phi}
\def\Pp{\phi^{\prime}}
\def\Ppp{\phi^{\prime\prime}}

\def\e{\epsilon}
\def\td{\tilde{d}}
\begin{document} 
\begin{flushright}
DPNU-01-28\\ hep-th/0109140
\end{flushright}

\vspace{10mm}

\begin{center}
{\Large \bf 
 Various Types of Five-Dimensional Warp Factor
 and Effective Planck Scale}

\vspace{20mm}
 Masato ITO 
 \footnote{E-mail address: mito@eken.phys.nagoya-u.ac.jp}
\end{center}

\begin{center}
{
\it 
{}Department of Physics, Nagoya University, Nagoya, 
JAPAN 464-8602}
\end{center}

\vspace{25mm}

\begin{abstract}
 Based on the assumption that the warp factor of
 four-dimensional spacetime and the one of fifth dimension are tied
 through a parameter $\alpha$, 
 we consider five-dimensional gravity with a $3$-brane coupled to a
 bulk scalar field.
 For arbitrary value of $\alpha$, the form of the 
 warp factor is implicitly determined by hypergeometric function.
 Concretely, we show that the warp factor becomes explicit form for
 appropriate value of $\alpha$,
 and study the relation between four-dimensional effective Planck scale and 
 the brane tension.
 This setup allows the possibility of extending the diversity
 of brane world. 
\end{abstract} 
\newpage 
%
%
 \section{Introduction}
 
 After Hor$\check{\rm a}$va-Witten model in eleven dimensional SUGRA 
 theory with two branes on $S^{1}/Z_{2}$ were proposed
 \cite{Horava:1996qa,Horava:1996ma}, the path for the study of brane
 world scenario opened rapidly.
 Especially, in the framework of five-dimensional theory with warp
 factor \cite{Randall:1999ee,Randall:1999vf,Rubakov:1983bz,Ito:2001fd} 
 (warped metric function) which is exponential damping function
 of fifth dimension compactified on  $S^{1}/Z_{2}$ orbifold, 
 Randall and Sundrum suggested that the origin of the hierarchy problem 
 comes from the warped geometry of the extra dimension.
 This scenario appeals to the possibility of an extra dimension
 limited by two $3$-branes with tensions of opposite sign
 \cite{Randall:1999ee}.
 Moreover, if fifth dimension is not compactified, massless graviton
 is localized on the brane and this setup has usual four-dimensional
 gravity law at large distance \cite{Randall:1999vf}.
 
 The Randall-Sundrum scenario including a bulk scalar field
 has been proposed \cite{Collins:2001ni,Collins:2001ed}.
 Identifying the scalar field with dilaton field, the scenario is
 inspired by string theory \cite{Gherghetta:2001iv}, sequentially,
 the solutions to the cosmological constant
 problem are very actively being investigated
 \cite{Csaki:2000wz,Binetruy:2000wn}. 
 In recent papers, the self-tuning mechanism of the cosmological constant
 has been suggested \cite{Kachru:2000xs,Kachru:2000hf}.
 The self-tuning idea allows a flat space solution without fine-tuning 
 between input parameters in lagrangian.
 Moreover, the study of cosmology in the framework of Randall-Sundrum
 scenario is widely performed,
 for instance, inflation in five-dimensional universe
 \cite{Nihei:1999mt,Kaloper:1999sm},
 a cosmology with radion stabilization \cite{Csaki:2000mp}
 and with a localized graviton \cite{Csaki:2000fc}.
 Thus it is expected that the Randall-Sundrum scenario has the 
 possibilities of making a breakthrough in particle physics or
 cosmology, and we are very interested in the form of the warp factor
 because the setup of the scenario is based on the ansatz for
 metric.

 In this paper, we point out that the form of the warp factor
 has various types in the framework of five-dimensional gravity with
 a $3$-brane coupled to a bulk scalar field.
 As for metric, we make the assumption that the warp factor
 of four-dimensional spacetime and the one of fifth
 dimension are related through a parameter $\alpha$ defined
 in ref.\cite{Ito:2001jk}.
 This metric corresponds to an extension of metric in the original
 Randall-Sundrum model and a new path will open the study of
 warp factor which is solution of Einstein equations.
 This is because the metric taken here corresponds to the most general metric
 appearing to SUGRA theories.
 In ref. \cite{Gherghetta:2001iv}, it was shown that 
 Randall-Sundrum brane-worlds arise from extremal D-brane configuration.
 Moreover, from a field theory point of view these brane-worlds consist
 of a warped geometry with a bulk scalar field, and the dual theories
 turn out to be non-conformal.
 To study the connection between Randall-Sundrum model and SUGRA theories,
 it is important to investigate various types of warp factor in the
 setup with a bulk scalar fields.
 As a result of investigation, 
 for arbitrary value of $\alpha$, the form of warp factor with
 dependence of fifth dimension is implicitly determined by
 hypergeometric function \cite{Ito:2001jk}.
 Moreover, for particular value of $\alpha$, 
 we show that the explicit form of the warp factor and scalar field
 have variety.
 Under the assumption of infinite fifth dimension,
 it is important to study whether effective Planck scale is finite or not.

 This paper is organized as follows. In section 2, we describe an action
 of the model and an ansatz for metric considered here.
 Solving the equations of motion, we obtain general form of warp factor
 controlled by both a parameter $\alpha$ and the sign of bulk cosmological 
 constant.
 In section 3, we explore the warp factor for four cases,
 where $\alpha=0,8,12$ and vanishing cosmological constant are
 described, separately.  
 For each case, we compute the effective Planck scale, brane tension
 and the relation between these.
 A conclusion is given in final section. 
%
%
 \section{The Model}

 We consider the following action
 \begin{eqnarray}
  S&=&\int d^{5}x\;\sqrt{-G}
                \left\{
                 \frac{1}{2\k^{2}_{5}}\R
                -\frac{1}{2}\left(\nabla\phi\right)^{2}-\L
                \right\}
   +\int d^{4}x\;\sqrt{-g}\;
      \left\{\;-f(\P)\;\right\}\,,
   \label{eq1}
 \end{eqnarray}
 where $\L$ is the cosmological constant in the bulk and 
 $1/\kappa^{2}_{5}=M^{3}_{\ast}$, $M_{\ast}$ is the
 fundamental scale of five-dimensional theory.
 Here $G$ is the five-dimensional metric and $g$ is the induced 
 four-dimensional metric on the brane which is located at $y=0$, 
 where $y$ is the coordinate of fifth dimension.
 In the second term, $f(\P)$ represents a brane tension
 coupled to a scalar field.
 We take the ansatz for metric \cite{Ito:2001jk}
 \begin{eqnarray}
  ds^{2}&=&e^{2A(y)}g_{\mu\nu}dx^{\mu}dx^{\nu}+e^{2\alpha A(y)}dy^{2}
  \nonumber\\
  &\equiv&G_{MN}dx^{M}dx^{N}\label{eq2}\,,
 \end{eqnarray}
 where $g_{\mu\nu}=diag(-,+,+,+)$.
 Note that the warp factor with $y$-dependence of 
 four-dimensional spacetime and the one of fifth dimension
 part are tied through a parameter $\alpha$.

 With the metric, the equations of motion are given by  
 \begin{eqnarray}
 &&\App + (2-\alpha)\left(\Ap\right)^{2}=
 -\frac{\k^{2}_{5}}{3}\left\{
  \frac{1}{2}(\Pp)^{2}+e^{2\alpha A}\L\right\}
 -\frac{\k^{2}_{5}}{3}e^{\alpha A}f(\P)\d (y)
  \label{eq3}\,,\\
 &&\left(\Ap\right)^{2}=
 \frac{1}{12}\k^{2}_{5}\left(\Pp\right)^{2}-\frac{\k^{2}_{5}}{6}
 e^{2\alpha A}\L
 \label{eq4}\,,\\
 &&(4-\alpha)\Ap\Pp+\Ppp
 =e^{\alpha A}\frac{\p f}{\p\P}\d(y)\,,\label{eq5}
 \end{eqnarray}
 where the prime represents the derivative with respect to the $y$.
 From Eqs.(\ref{eq4}) and (\ref{eq5}), the equations in the bulk
 are expressed as
 \begin{eqnarray}
  \Pp &=&c\; e^{(\alpha-4)A}\,.\label{eq6}\\
  \Ap &=&\e\frac{\sqrt{3}}{6}\k_{5}\left|c\right|
      e^{(\alpha-4)A}\sqrt{1-\frac{2\L}{c^{2}}e^{8A}}\,,
  \label{eq7}
 \end{eqnarray}
 Hence $c$ is the integration constant and $\e=\pm$, where the sign 
 $\e$ determines the branch of the square root.

 From Eqs.(\ref{eq6}) and (\ref{eq7}),
 we obtain the solution for a scalar field
 \begin{eqnarray}
  \P(y)=\e\frac{c}{|c|}\frac{2\sqrt{3}}{\k_{5}}A(y)+k\label{eq8}
 \end{eqnarray}
 for $\L=0$, and
  \begin{eqnarray}
  \P(y)=-\e\frac{c}{|c|}\frac{\sqrt{3}}{2\k_{5}}\tanh^{-1}
         \sqrt{1-\frac{2\L}{c^{2}}e^{8A(y)}}+l
  \label{eq9}
 \end{eqnarray}
 for $\L\neq 0$. Here $k$ and $l$ are the integration constants.
 Consequently, $A(y)$ is expressed as \cite{Ito:2001jk}
 \begin{eqnarray}
  e^{(4-\alpha)A}\;
 {}_{2}F_{1}\left(\;\frac{1}{2}\,,\frac{4-\alpha}{8}\,;
 \frac{12-\alpha}{8}\,;\,\frac{2\L}{c^{2}}e^{8A}\;\right)
 =\e\frac{\sqrt{3}}{6}\k_{5}\left|c\right|(4-\alpha)
 \left(y+d\right)\,,
 \label{eq10}
 \end{eqnarray}
 for $\alpha\neq 4,12$.
 Here $d$ is the integration constant and we can take normalization
 condition $A(0)=0$.
 In the case of $\alpha=4,12$, since we cannot use the integral
 representation of hypergeometric function due to vanishing of second or
 third argument, $A(y)$ can be directly obtained by the integration of
 Eq.(\ref{eq7}).
 The case of $\alpha=4$ was investigated in ref.\cite{Ito:2001jk}, 
 we obtained the warp factor
 without having a singularity and showed that the bulk 
 cosmological constant is bounded by both finite effective Planck scale
 and five-dimensional fundamental scale.
 Mathematically, adopting specific value of $\alpha$,
 the hypergeometric function can be described explicitly in terms
 of elementary function\footnote{
 Examples of the hypergeometric function represented by elementary
 function :\\ $\displaystyle 
 {}_{2}F_{1}\left(\frac{1}{2},\frac{1}{2};\frac{3}{2};-z^2\right)=
 \frac{\sinh^{-1}z}{z}\;\,,\;
 {}_{2}F_{1}\left(\frac{1}{2},-\frac{1}{2};\frac{1}{2};z^2\right)=
 \cos\left(\sin^{-1}z\right)\;\,,\;
 {}_{2}F_{1}\left(a,b;c;0\right)=1$}.
 As mentioned later, we describe concrete form of warp factor.

 Due to delta function in Eqs.(\ref{eq3}) and (\ref{eq5}),
 the jump conditions with respect to the first derivative of $A$ and
 $\P$ can be obtained.
 We get the following results for the jump conditions using
 Eqs.(\ref{eq3}) and (\ref{eq5})
 \begin{eqnarray}
  \e_{+}\frac{1}{\td_{+}}-\e_{-}\frac{1}{\td_{-}}
  &=&\frac{\p f}{\p\P}(\P(0))\,,\nonumber\\
  \e_{+}\sqrt{\frac{1}{\td^{2}_{+}}-2\L}-
  \e_{-}\sqrt{\frac{1}{\td^{2}_{-}}-2\L}&=&
  -\frac{2}{\sqrt{3}}\k_{5}f(\P(0))\,,\label{eq11}
 \end{eqnarray}
 where 
 \begin{eqnarray}
  \frac{1}{\td_{\pm}}=\frac{6}{\sqrt{3}\k_{5}(4-\alpha)}\;
  {}_{2}F_{1}\left(\;\frac{1}{2}\,,\frac{4-\alpha}{8}\,;
 \frac{12-\alpha}{8}\,;\,\frac{2\L}{c^{2}_{\pm}}\;\right)
 \frac{1}{d_{\pm}}\,.\label{eq12}
 \end{eqnarray} 
 We denote the sign and the integration constants for the positive
 region $(y>0)$ by $\e_{+}\,,c_{+}\,,d_{+}$ and those for the negative 
 region $(y<0)$ by $\e_{-}\,,c_{-}\,,d_{-}$.
 Using the above equations, we can obtain the brane tension at $y=0$.

 Furthermore, if the fifth dimension is assumed to have $Z_{2}$
 symmetry, we can impose the condition
 $\e_{+}=-\e_{-}$ and $c_{+}=-c_{-}=c$.
 Integrating out the fifth dimension, the four-dimensional
 effective Planck scale is given by
 \begin{eqnarray}
 M^{2}_{\rm Pl}&=&\frac{1}{\k^{2}_{5}}
 \left|\;\int\;dy\;e^{(2+\alpha)A(y)}\;\right|\nonumber\\
 &=&\frac{\sqrt{3}}{2\k^{3}_{5}|c|}
 \left|\;\left[\;
 t^{6}\;{}_{2}F_{1}\left(\;\frac{3}{4}\,,\frac{1}{2}\,;
 \frac{7}{4}\,;\frac{2\L}{c^{2}}t^{8}\;\right)\;
 \right]^{t=e^{A_{+}(r_{c})}}_{t=1}
 \;\right|\,,\label{eq13}
 \end{eqnarray}
 where $A_{+}$ corresponds to the function
 for positive region $(y>0)$.
 From the second line in Eq.(\ref{eq13}), the effective Planck scale
 is represented not in terms of $\alpha$ but
 in terms of $A$.
 Thus, the finite of effective Planck scale is determined by both the
 sign of $\L$ and
 the value of warped metric function at $y=r_{c}$.
 In this setup, since we consider the case of infinite fifth dimension,
 the limit of $r_{c}\rightarrow\infty$ must be taken.
 Below, we investigate the form of warp factor and
 the finite of effective Planck scale for four specific value of $\alpha$.

%
 \section{Solutions}

 We study the two cases of $\alpha=0,8$ where the hypergeometric
 function in Eq.(\ref{eq10}) is represented by elementary function, 
 and the case of $\alpha=12$
 where hypergeometric function cannot be used, and the case of 
 vanishing bulk cosmological constant, separately. 
 Our setup is based on the assumption that the
 fifth dimension is not compactified with $Z_{2}$ symmetry.  
 For four cases, we obtain the warp factor and estimate the brane
 tension $(\; V=f(\P(0))\;)$ and
 four-dimensional effective Planck scale $M_{\rm Pl}$.

 \subsection{Case I}

 For $\alpha=0$ and $\L<0$, from footnote $\dagger$, we can obtain 
 the warp factor as follows
 \begin{eqnarray}
  e^{4A_{\pm}}=\frac{|c|}{\sqrt{2|\L|}}
  \sinh\left(\;\pm\frac{2\sqrt{6|\L|}}{3}\k_{5}\;(\;y\pm d\;)\;\right)\,,
  \label{eq14}
 \end{eqnarray}
 where plus ( minus ) corresponds to the function for $y>0$
 ( $y<0$ ) region.
 The condition $A(0)=0$ yields
 \begin{eqnarray}
  \sinh\frac{2\sqrt{6|\L|}}{3}\k_{5}d
 =\frac{\sqrt{2|\L|}}{|c|}\,.\label{eq15}
 \end{eqnarray}
 Moreover, the jump condition at $y=0$ leads to the relation between
 the brane tension $V$ and the bulk cosmological constant $\L$
 \begin{eqnarray}
  V=-\frac{\sqrt{3(c^{2}+2|\L|)}}{\k_{5}}\,.
  \label{eq16}
 \end{eqnarray} 
 From Eq.(\ref{eq13}), we get
 \begin{eqnarray}
  M^{2}_{\rm Pl}=\frac{2\sqrt{3}}{3\k^{3}_{5}|c|}
  \left|\;\left(\;\frac{|c|}{\sqrt{2|\L|}}\;\right)^{\frac{3}{2}}
  s^{\frac{3}{2}}\;F\left(-s^{2}\right)
  -F\left(-\frac{2|\L|}{c^{2}}\right)
  \;\right|\,,\label{eq17}
 \end{eqnarray}
 where ${}_{2}F_{1}\left(\frac{3}{4},\frac{1}{2};\frac{7}{4};z\right)
 \equiv F(z)$ and we defined
 \begin{eqnarray}
  s=\sinh\frac{2\sqrt{6|\L|}}{3}\k_{5}(r_{c}+d)\,.
 \label{eq18}
 \end{eqnarray}
 Since infinite fifth dimension corresponds to the limit of 
 $s\rightarrow\infty$, Eq.(\ref{eq17}) leads infinite effective Planck
 scale.
 Namely, in order to obtain the finite of effective Planck scale,
 it is necessary to consider the setup including another $3$-brane at
 $y=r_{c}$ with finite fifth dimension. 
 Furthermore, from Eq.(\ref{eq9}), the scalar field is infinite
 everywhere due to negative cosmological constant in the bulk.
 Therefore, this case is excluded.
  The case of $\alpha=0,\L>0$ had been already investigated 
 \cite{Collins:2001ni,Collins:2001ed}. 

 \subsection{Case II}

 For $\alpha=8$ and $\L>0$, from footnote $\dagger$ and Eq.(\ref{eq10}),
 we have
 \begin{eqnarray}
  e^{8A_{\pm}}=\left[\;\frac{2\L}{c^{2}}+\frac{4}{3}k^{2}_{5}c^{2}
  (\;y\pm d\;)^{2}\;\right]^{-1}\,.\label{eq19}
 \end{eqnarray}
 Imposing the condition $A(0)=0$,
 we obtain the condition for the integration constants
 \begin{eqnarray}
  \frac{2\L}{c^{2}}+\frac{4}{3}k^{2}_{5}d^{2}c^{2}=1\,,
  \label{eq20}
 \end{eqnarray}
 and the jump condition at $y=0$ yields the brane tension
 \begin{eqnarray} 
 V=2c^{2}d\,.\label{eq21}
 \end{eqnarray}
 From the above equations,
 the integration constant $d$ can be eliminated as follows
 \begin{eqnarray}
  2\L+\frac{1}{3}\k^{2}_{5}V^{2}=c^{2}\,.
  \label{eq22}
 \end{eqnarray}
 Thus, the brane tension is given by
 \begin{eqnarray}
  V=\pm\frac{\sqrt{3(c^{2}-2\L)}}{\k_{5}}\,,
  \label{eq23}
 \end{eqnarray}
 therefore, the cosmological constant in the bulk is bounded
 \begin{eqnarray}
  0<\L\leq\frac{c^{2}}{2}\label{eq24}\,.
 \end{eqnarray}
 In this case, effective Planck scale is expressed as
 \begin{eqnarray}
  M^{2}_{\rm Pl}=\frac{2\sqrt{3}}{3\k^{2}_{5}|c|}
  \left|\;
  u^{-\frac{3}{4}}F\left(\;\frac{2\L}{c^{2}u}\;\right)
  -F\left(\;\frac{2\L}{c^{2}}\;\right)
  \;\right|\,,\label{eq25}
 \end{eqnarray}
 where we define $u$ by
 \begin{eqnarray}
  u=\frac{2\L}{c^{2}}+\frac{4}{3}\k^{2}_{5}c^{2}(r_{c}+d)^{2}\,.
  \label{eq26}
 \end{eqnarray}
 Infinite fifth dimension $(u\rightarrow\infty)$ yields finite
 effective Planck scale 
 \begin{eqnarray}
  M^{2}_{\rm Pl}=\frac{2\sqrt{3}}{3\k^{3}_{5}|c|}
  F\left(\;\frac{2\L}{c^{2}}\;\right)\,.
  \label{eq27}
 \end{eqnarray}
 Using Eq.(\ref{eq24}) and the fact that 
 $0<F(z)\leq \frac{\Gamma(\frac{1}{2})
 \Gamma(\frac{7}{4})}{\Gamma(\frac{5}{4})}\sim 1.8$ for $0<z\leq 1$,
 \begin{eqnarray}
  M^{2}_{\rm Pl}\sim\frac{M^{\frac{9}{2}}}{|c|}\,.
  \label{eq28}
 \end{eqnarray}
 From Eq.(\ref{eq23}), the bulk cosmological constant becomes
 \begin{eqnarray}
  2\L\sim\frac{M^{9}_{\ast}}{M^{4}_{\rm Pl}}-\frac{V^{2}}{3M^{3}_{\ast}}
  \label{eq29}\,,
 \end{eqnarray}
 and the condition $\L>0$ leads to the following inequality
 \begin{eqnarray}
  M_{\ast}>V^{\frac{1}{6}}M^{\frac{1}{3}}_{\rm Pl}\,.
  \label{eq30}
 \end{eqnarray}  
 Thus, effective Planck scale is finite and the lower bound of 
 five-dimensional fundamental scale is
 described in terms of both the brane tension and the effective
 Planck scale.
 Eqs.(\ref{eq29}) and (\ref{eq30}) are as same as
 the case of $\alpha=4$ in ref.\cite{Ito:2001jk}.
 Moreover, using Eqs.(\ref{eq9}) and (\ref{eq19}), the scalar field in
 the bulk is described in terms of $y$.

 \subsection{Case III}

 For $\alpha=12$, since third argument of the hypergeometric function
 in Eq.(\ref{eq10}) is zero, we directly integrate Eq.(\ref{eq7}).
 The solution is given by
 \begin{eqnarray}
  \tanh^{-1}\sqrt{1-\frac{2\L}{c^{2}}e^{8A_{\pm}}}
 +\frac{c^{2}}{2\L}e^{-8A_{\pm}}\sqrt{1-\frac{2\L}{c^{2}}e^{8A_{\pm}}}
 =\pm\frac{2\sqrt{3}k_{5}|c|^{3}}{3\L}(y\pm d)\,,
 \label{eq31}
 \end{eqnarray}
 using $A(0)=0$, we get
 \begin{eqnarray}
  \tanh^{-1}\sqrt{1-\frac{2\L}{c^{2}}}
 +\frac{c^{2}}{2\L}\sqrt{1-\frac{2\L}{c^{2}}}
 =\frac{2\sqrt{3}k_{5}|c|^{3}}{3\L}d\,.\label{eq32}
 \end{eqnarray} 
 Note that this solution only makes sense when the argument of
 inverse function of hyperbolic tangent is smaller than unity, namely,
 \begin{eqnarray}
  0<\L\leq \frac{c^{2}}{2}\,.\label{eq33}
 \end{eqnarray}
 From the above equations, the brane tension is expressed as
 \begin{eqnarray}
  V=\frac{\sqrt{3(c^{2}-2\L)}}{k_{5}}\,.\label{eq34}
 \end{eqnarray}
 In infinite fifth dimension, from Eq.(\ref{eq31}), 
 we get $e^{A(\infty)}=0$.
 Taking account of the range of $\L$ in Eq.(\ref{eq33}), 
 the effective Planck scale is expressed as
 \begin{eqnarray}
  M^{2}_{\rm Pl}\sim\frac{1}{\k^{3}_{5}|c|}
  =\frac{M^{\frac{9}{2}}_{\ast}}{|c|}\label{eq35}\,.
 \end{eqnarray}
 This result is as same as the case of $\alpha=8$.
 Namely, the relation between effective Planck scale and brane tension
 is Eq.(\ref{eq29}).
 Using Eqs.(\ref{eq9}) and (\ref{eq31}), the relation between 
 bulk scalar field and $y$ can be obtained.
 
 \subsection{Case IV}

 For $\L=0$ and $\alpha\neq 4$, we obtain the following warp factor
 \begin{eqnarray}
  A_{\pm}(y)=\frac{1}{4-\alpha}\ln
  \left[\;\pm\frac{|4-\alpha|}{2\sqrt{3}}\k_{5}|c|(\;y\pm d\;)\;\right]
  \label{eq36}\,,
 \end{eqnarray}
 therefore, using the condition $A(0)=0$,
 the brane tension is expressed as
 \begin{eqnarray}
  V=\pm\frac{\sqrt{3}|c|}{\k_{5}}\,,
  \label{eq37}
 \end{eqnarray}
 where the sign $+$ ($-$) corresponds to the case of $\alpha>4$
 ($\alpha<4$). 

 The effective Planck scale is 
 \begin{eqnarray}
  M^{2}_{\rm Pl}=\frac{\sqrt{3}}{2\k^{3}_{5}|c|}
  \left|\;v^{\frac{6}{4-\alpha}}-1\;\right|\,,
  \label{eq38}
 \end{eqnarray}
 where $\displaystyle
 v\equiv \frac{\k_{5}|c|}{2\sqrt{3}}|4-\alpha|(r_{c}+d)$.
 In infinite fifth dimension ($v\rightarrow\infty$),
 the finite of effective Planck scale is determined
 by the sign of exponent of $v$ in Eq.(\ref{eq38}).
 Namely, for $\alpha<4$,
 this case must be excluded because the effective Planck scale is 
 infinite.
 However, for $\alpha>4$, the effective Planck scale is finite
 \begin{eqnarray}
  M^{2}_{\rm Pl}\sim\frac{M^{6}_{\ast}}{V}\,,
  \label{eq39}
 \end{eqnarray}
 where we used the result of Eq.(\ref{eq37}).
 Obviously, vanishing brane tension leads infinite effective Planck scale.
 If the brane tension is TeV scale, five-dimensional fundamental scale
 becomes
 \begin{eqnarray}
  M_{\ast}\sim 10^{8}\;{\rm GeV}\label{eq40}
 \end{eqnarray}
 In addition, using Eqs.(\ref{eq8}) and (\ref{eq36}), 
 the scalar field in the bulk is represented in terms of
 logarithmic function.
 This behavior is similar to the result of ref.\cite{Kachru:2000hf}.

%
 \section{Conclusion}

 We have presented various patterns of warp factor in brane world
 with a bulk scalar field.
 Under the assumption that the warp factor $e^{B(y)}$ of fifth dimension
 is related with the one $e^{A(y)}$ of four-dimensional spacetime:
 $B=\alpha A$, we were able to find solutions of equations of motion
 for specific value of $\alpha$.
 The metric taken here corresponds to the metric appearing in SUGRA
 theories to be low energy effective theories of D-brane configurations.
 This setup corresponds to an extension of Randall-Sundrum type,
 and it may be possible to clear the conjecture that Randall-Sundrum model
 arises from the underlying theories (string or M-theory)

 Based on a model with infinite fifth dimension,
 for four cases where $\alpha=0,8,12$ as well as vanishing bulk
 cosmological constant, we obtained the warp factor and brane tension,
 and investigated whether effective Planck scale is finite or not. 
 The case of $\alpha=0,\L<0$ is excluded because of infinite effective
 Planck scale, in addition, bulk scalar field is infinite everywhere.
 In the case of $\alpha=8,\L>0$, warp factor has fractional form, and
 finite effective Planck scale is obtained.
 It is shown that bulk cosmological constant is bounded by integration
 constant, and five-dimensional fundamental scale has lower bound.
 In the case of $\alpha=12$, finite Planck scale,
 bulk cosmological constant is bounded,
 and brane tension becomes positive.
 In the case of $\alpha\neq 4,\L=0$, both warp factor and scalar field have
 logarithmic form.
 The case of $\alpha<4$ is excluded due to infinite effective Planck.
 For $\alpha>4$, we pointed out that effective Planck scale is finite,
 however, it is infinite if brane tension is zero.
 Thus, we explored the diversity of the five-dimensional warped
 geometry.
 
 There are articles which we discuss in future.
 It is necessary to study the massless gravitational fluctuations about
 our classical solution obtained here.
 Moreover, we are interested in cosmological evolution for
 each type of warp factor.
 This setup can be possibly applied to several scenarios.
 We will turn to these works elsewhere. 
%
 
\end{document}